# Distribution of Localized States from Fine Analysis of Electron Spin Resonance Spectra in Organic Transistors

Hiroyuki Matsui (松井 弘之)

*Photonics Research Institute (PRI), AIST, Tsukuba 305-8562, Japan*

*Department of Advanced Materials Science, The University of Tokyo, Kashiwa, Chiba 277-8561, Japan*

Andrei S. Mishchenko

*Cross-Correlated Materials Research Group (CMRG), ASI, RIKEN, Wako 351-0198, Japan*

*RRC Kurchatov Institute, 123182, Moscow, Russia*

Tatsuo Hasegawa (長谷川 達生)

*Photonics Research Institute (PRI), AIST, Tsukuba 305-8562, Japan*

We developed a novel method for obtaining the distribution of trapped carriers over their degree of localization in organic transistors, based on the fine analysis of electron spin resonance spectra at low enough temperatures where all carriers are localized. To apply the method to pentacene thin-film transistors, we proved through continuous wave saturation experiments that all carriers are localized at below 50 K. We analyzed the spectra at 20 K and found that the major groups of traps comprise localized states having wave functions spanning around 1.5 and 5 molecules and a continuous distribution of states with spatial extent in the range between 6 and 20 molecules.

Electron spin resonance (ESR) offers a unique microscopic probe to detect the transition from charge localization to delocalization in solids. A monumental achievement in this field was establishing the Mott-Anderson transition from insulator to metal in phosphorous-doped silicon crystals in the late 1950s [1,2]; a drastic change was observed from hyperfine structures at low doping concentration to a single narrow peak at higher doping concentration as associated with impurity band formation. Analysis of the hyperfine structures also provided specific information about the number of acceptor atoms, $^{31}$P ($I = 1/2$), over which each localized wave function extends [3]. Recently, the field-induced ESR (FESR) technique was successfully utilized to demonstrate the delocalized (or mobile) nature of the carriers that accumulate through applied gate voltage in organic transistors. Typical pentacene thin-film transistors (TFTs) [4] and rubrene single-crystal transistors [5] exhibit fairly sharp FESR spectra whose single-Lorentzian linewidth presents motional narrowing effects with increase of the temperature. In case of the pentacene TFTs, the feature is well consistent with the thermally-activated multiple trap-and-release (MTR) transport with the activation energy of about 10 meV in the high temperature range. In contrast, the narrowing effect is not observed below around 50 K where the variation of the linewidth becomes saturated, while the transition from the delocalization to the localization is not at all clear in the FESR spectra. The most puzzling issue is that the FESR spectra remain similar to the single Lorentzian curves rather than Gaussian [6,7] down to low enough temperature despite the expected charge localization.

In this Letter, we first demonstrate by continuous wave (cw) saturation experiments that the FESR spectra at temperature lower than 50 K are understood in terms of the charge localization and that the hyperfine mechanism dominates the lineshape in the low temperature range. Then we suggest a novel unbiased method for fine analysis of the low-temperature FESR spectra, based on simple assumptions that the localized states are composed of several kinds of trap states with different spatial extensions and that the each kind of trap state should afford single Gaussian spectrum due to the multiple hyperfine interactions of nuclear spins within the molecules. The method enables us to reveal the distribution of trap states over the spatial extent of the wave functions. The analysis identified three major groups of trap levels. The first two have localized wave functions spanning about 1.5 and 5 molecules, respectively, and the third one has a continuous distribution ranging from 6 to 20 molecules. We relate the spatial extent to the binding energy using Holstein model for trapped polaron [8].

For the FESR experiments, we fabricated pentacene TFTs with a bottom-gate, top-contact geometry, where Parylene C was used as the gate dielectric layer (thickness: 800 nm, capacitance: 4.5 nF/cm$^2$) [4]. In the FESR measurements at low temperature, we first applied gate voltages at room temperature (with source and drain shorted) and then cooled down the device to the set temperature, in order to avoid the delay of the charge accumulation. In the cw saturation experiments, we measured the integrated intensity $I$ and the linewidth $\Delta B_{1/2}$ (half width at half maximum) of FESR spectra as a function of the microwave power $P$ at the gate voltage $V_G = -200$ V. We observed the intensity saturation with increasing $P$ at the respective temperatures, as shown in Fig. 1(a). All plots followed typical saturation curves as $I \propto [P/(P+P_S)]^{1/2}$ [9], while saturation started at different microwave power $P_S$ at different temperatures. $P_S$ increases with increasing temperature, as shown in the inset of Fig. 1(a), which implies that the excited spin states relax more rapidly at higher temperatures. Figure 1(b) shows the variation in linewidth under saturation. The plot allows us to estimate the homogeneity of the FESR spectra: the broadening under saturation indicates that the spectra are more or less homogeneous due to the motional narrowing. The FESR spectra are fully homogeneous at 283 K, since the plot is closest to the ideal curves for homogeneous broadening: $\Delta B_{1/2} \propto [(P+P_S)/P_S]^{1/2}$ (Curve A) [9]. In sharp contrast, the linewidth does not show any broadening under saturation below 50 K. This indicates that the spectra are inhomogeneous and that the motional narrowing is not effective in the low-temperature range. From these results, we concluded that the average trap residence time, $\tau_C$, exceeds the upper limit for motional narrowing; $\tau_C > (\gamma \Delta B_{\text{inhomo}})^{-1}$ ($\gamma$ is the electron gyromagnetic ratio, and $\Delta B_{\text{inhomo}}$ is the inhomogeneous local magnetic field) at around 50 K [10]. This picture correlates well with the temperature dependence of $\Delta B_{1/2}$ in Fig. 2; the $\Delta B_{1/2}$ becomes saturated below 50 K. We also estimated the spin-lattice relaxation time as shown in Fig. 2, indicating that the contribution of spin-lattice relaxation is negligible at low temperatures. All of these results afford clear evidence that the FESR spectra below 50 K are derived from frozen carriers at the trap sites, where the hyperfine mechanism dominates the spectral features.

Here, we focus on the low-temperature FESR spectra. Unlike the powder-pattern lineshape, we observed highly symmetric FESR spectra for all magnetic field directions and at any temperature, irrespective of the anisotropic molecular $g$ tensor in polycrystalline forms. The feature is partly associated with the high $c$-axis orientation of the polycrystalline films [11]. It is also considered from the symmetric nature at the field direction within the film plane that the anisotropic molecular $g$ tensor should be averaged out over several molecules with different molecular orientations. Actually, the estimated FESR linewidth at 20 K is about 180 µT, which is about one third of that for isolated cations in solution [13] (see Fig. 2). These features afford unambiguous evidence that the wave functions of frozen (or trapped) carriers extends over several molecules. If we assume that the trap state extends over $N$ molecules, the hyperfine interactions should afford single Gaussian spectrum with smaller linewidth than that of isolated molecule by a factor of $N^{-1/2}$ [6,12]. Since the observed FESR spectrum is not a single Gaussian, it is concluded that the spectrum originates from various kinds of trap states with different spatial extensions. To show this, the first-derivative FESR spectrum $S(B)$ is represented by the sum of multiple Gaussians with different linewidths:

$$S(B) = \int_1^{+\infty} D(N) \frac{\partial G(B,N)}{\partial B} dN, \qquad (1)$$

$$G(B,N) = \sqrt{\frac{N}{2\pi\sigma_0^2}} \exp\left[-\frac{N(B-B_0)^2}{2\sigma_0^2}\right]. \qquad (2)$$

Here, $D(N)$ is the distribution function for $N$, $G(B,N)$ the normalized Gaussian with the linewidth $\sigma_0 N^{-1/2}$, $\sigma_0$ the linewidth at $N = 1$, and $B_0$ the resonance magnetic field. In order to secure the generality, we redefine the extension of trap states with the probability $\{p_i\}$ on molecular site $i$, $\Sigma_i\, p_i = 1$. Then the scaling factor becomes $(\Sigma_i\, p_i^2)^{1/2}$ so that the effective spatial extent is represented as $N = (\Sigma_i\, p_i^2)^{-1}$, which is not limited to the integer number. By analyzing the experimental $S(B)$ using stochastic optimization method [14], we obtained the unknown distribution $D(N)$. The method does not assume any specific function forms for $D(N)$ and, hence, the procedure does not require any *a priori* assumptions.

For the spectral analysis of localized spin states, we carried out high-precision FESR measurements at 20 K, the spectrum of which is shown in Fig. 3 (a). A large number of field-induced spins were accumulated by stacking ten sheets of TFTs. The static magnetic field was set perpendicular to the film plane, which reduces the anisotropic effect of $g$ values. The residuals between the experimental and the calculated spectra were composed of white noise in contrast to the correlated fluctuation in case of the Lorentzian fitting (see Fig. 3(b)). The resultant distribution $D(N)$ presented good reproducibility, and is composed of two discrete peaks (A and B) at $N = 1.5 \pm 0.13$ ($0.9\times10^{12}$ cm$^{-2}$) and $5 \pm 0.19$ ($2.3\times10^{12}$ cm$^{-2}$) respectively, and a broad feature (C) at $N$ = 6–20 molecules ($2.3 \times 10^{12}$ cm$^{-2}$) (Fig. 4).

To relate the spatial extent $N$ to binding energy $E_B$ for the traps, we calculated these quantities using Holstein model for trapped polaron by diagrammatic Monte Carlo (DMC) method in direct space [8] where the polaron in a two-dimensional square lattice is coupled by the coupling constant $\gamma$ to intramolecular phonons with frequency $\omega_{ph}$. The polaron is trapped by a short-range on-site attractive potential $U$ and its coherent movement is described by the transfer integral $t$:

$$H = -U c_0^+ c_0 + \omega_{ph} \sum_i b_i^+ b_i - t \sum_{<i,j>} c_i^+ c_j - \gamma \sum_i (b_i^+ + b_i) c_i^+ c_i . \qquad (3)$$

Here, $c_i^+$ ($b_i^+$) is the creation operator for electron (phonon) at the $i$-th molecule. The parameters are $t$ = 0.1 eV [15], $\omega_{ph}$ = 0.1 eV and electron-phonon coupling constant $\lambda = (\gamma^2/4t\omega_{ph}) = 0.15$ [16]. We determined $N$ and $E_B$ as functions of $U$, and then obtained the relation between $E_B$ and $N$. Naturally, the larger attractive potential further localizes the wave function, and we found an empirical relationship:

$$N - 1 = \left[ E_B / \Delta(\lambda) \right]^{-1.1} , \qquad (4)$$

where $\Delta(0.15)$ = 75 meV. Equation (4) enabled us to obtain the distribution of trapped carriers on an energy scale. We found that the two discrete trap levels (A and B) peak at 140 ± 50 meV and 22 ± 3 meV, respectively, while the broad feature (C) is distributed between 5 and 15 meV, as presented in the inset of Fig. 4. The low-energy profile prompts the anticipation of tail states extending from just below the band edge, as has been discussed for amorphous semiconductors, while the states are partially occupied up to the Fermi level at around $E_B$ = 5 meV. These results are roughly consistent with the small activation energy of about 15 meV for the motional narrowing in Fig. 2, and also with the potential fluctuations by atomic-force-microscope potentiometry [17]. We note that the distribution $D(E_B)$ gives relatively correct position of the trap levels, although the absolute value of the binding energy is rather model-dependent.

Such weakly-localized in-gap states should take crucial roles in the intrinsic charge transport along semiconductor/gate dielectric interfaces in organic transistors. Actually, temperature-independent mobility is often observed in the devices with high mobility and highly-ordered molecular interfaces, which indicates that the Fermi energy is just below the band edge [18]. So far various experimental techniques have been utilized to investigate the interfacial trap density, such as deep-level transient spectroscopy (DLTS) [19], photocurrent yield [20], gate-bias stress [21], and thermally-stimulated current [22] experiments. However, the measurements are based on the charge transport that is considerably affected by the "extrinsic" potential barriers at grain boundaries and/or channel/electrode interfaces. In striking contrast, the present method has a crucial advantage in its ability to disclose the spatial and energy distribution of shallow traps down to a few meV, based on the unique microscopic probe using electronic spins. In addition, $g$ tensor can be utilized to identify the molecular species around the trap sites. For the obtained three types (A, B, and C) of trap states, the $g$ tensor should be common, considering the highly symmetric nature of the FESR spectra. It indicates that the trap states are extended over inherent pentacene molecules of regular orientation [4,6]. Among them, the deep discrete trap level (A) might be attributed to structural defects such as grain boundaries [23]. We can also consider that the shallow discrete level (B) and a broad feature (C) might be ascribed to small defects such as molecular sliding along the long axes of molecules [24], or disorder induced by random dipoles in the amorphous gate dielectrics [25,26]. Notice that the regular orientations of molecules are retained in the trap states as stated above. Although these assignments are rather speculative, we believe that further microscopic investigations based on this study should provide a distinct, comprehensive view on the weakly-localized in-gap states in organic transistors.

In summary, we developed a method for obtaining the distribution of trapped carriers over their degree of localization in organic TFTs, based on the fine analysis of FESR spectra. We first demonstrated through cw saturation experiments that the FESR spectra at $T < 50$ K are dominated by the hyperfine mechanisms of frozen carriers at trap sites. Analyzing the FESR spectrum at 20 K, we showed that the signal is split into the multiple Gaussian components, each of which corresponds to different spatial extensions. The major trap levels comprise localized wave functions spanning around 1.5 and 5 molecules and a broad feature at $N = 6$–20 molecules. Using the DMC calculations for Holstein model, we mapped the spatial extensions into the binding energies at 140 meV (A) and 22 meV (B), respectively, with the broad feature distributed at 5–15 meV (C). The filling of shallow traps up to 5 meV indicates that the Fermi energy is very close to the valence band edge. These shallow in-gap states should be important in understanding and improving the device performance of organic TFTs.

We are grateful to Prof. Kenji Mizoguchi and Prof. Bertram Batlogg for helpful discussions. This study is partly supported by JSPS, Grant-in-Aid for Scientific Research 09J06819. ASM acknowledges support provided by RFBR 07-02-0067a.

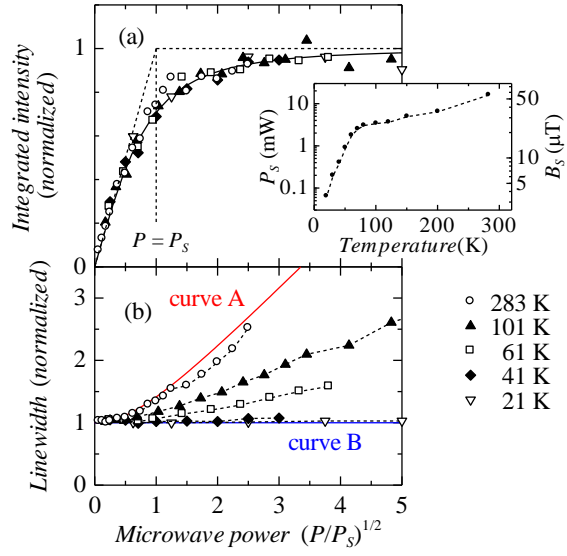

FIG. 1 (color online) Continuous wave saturation of (a) ESR intensity and (b) linewidth of pentacene TFTs at gate voltage of −200 V are plotted as a function of $(P/P_s)^{1/2}$, where $P$ is the incident microwave power and $P_s$ is the saturation microwave power. Curves A and B in (b) present the ideal lines for homogeneous and inhomogeneous cases. Inset shows the temperature dependence of $P_S$ and the corresponding magnetic field $B_S$.

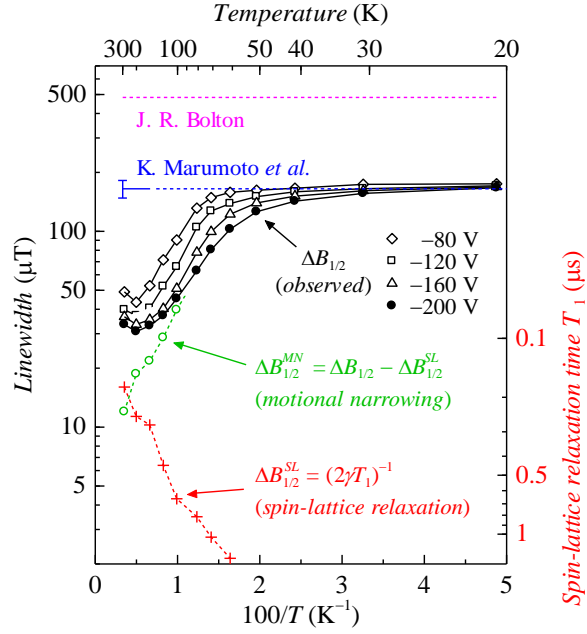

FIG. 2 (color online) Temperature dependence (Arrhenius plot) of ESR linewidths $\Delta B_{1/2}$ for pentacene TFTs at four different gate voltages. Reported values of linewidth for pentacene cation solution [13] and for low-mobility TFT [7] are also shown. $\Delta B_{1/2}^{MN}$ (green open circles) and $\Delta B_{1/2}^{SL}$ (red crosses) indicate the contribution of motional narrowing and spin-lattice relaxation effects, respectively, as obtained by cw saturation analysis of spin-lattice relaxation time $T_1$ ($T_1 = (2T_2\gamma^2 B_S^2)^{-1}$; $T_2$ is the spin-spin relaxation time related to the observed linewidth).

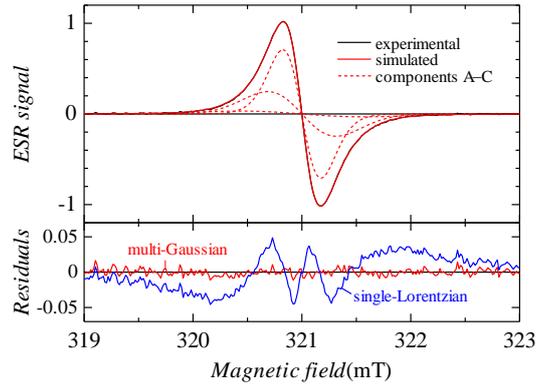

FIG. 3 (color online) (a) High-precision FESR spectrum of pentacene TFT measured at 20 K and at gate voltage of −200 V. Also shown is a simulated curve reproduced by multiple-Gaussian fitting, with three components indicated by dashed lines. (b) Residuals for the stochastic optimization analysis (red line) using the final distribution function $D(N)$ shown in Fig. 4. Residual for a simple Lorentzian fitting is also shown (blue line).

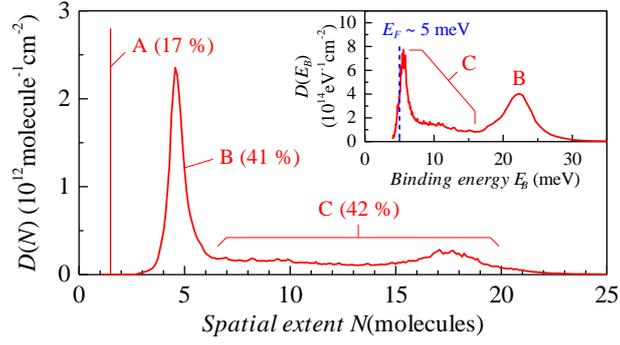

FIG. 4 (color online) (a) Distribution of trap states in pentacene TFTs are plotted against the spatial extent $N$ of the wave function, as obtained by stochastic optimization analysis of FESR spectrum at 20 K and at gate voltage of −200 V. Inset shows the distribution of trap states as a function of binding energy $E_B$ in the low-energy region, where the distribution as a function of $N$ is converted by approximation-free diagrammatic Monte Carlo method.